\documentclass[10pt,a4paper]{IEEEtran}
\usepackage{graphicx} 
\usepackage{amsmath,amssymb}
\usepackage{amsfonts}
\usepackage{mathrsfs}
\usepackage[noadjust]{cite}
\usepackage{xcolor}
\usepackage{comment}
\usepackage{enumitem}

\usepackage{color}

\newcommand{\ml}[1]{#1}

\newtheorem{theorem}{Theorem}
\newtheorem{lemma}[theorem]{Lemma}

\newtheorem{proposition}[theorem]{Proposition}

\newtheorem{definition}{Definition} 
\newtheorem{example}{Example}

\newtheorem{question}{Question}

\allowdisplaybreaks

\newcommand{\be}{\begin{equation}}
\newcommand{\ee}{\end{equation}}

\newcommand{\calC}{\mathcal{C}}

\newcommand{\calG}{\mathcal{G}}

\newcommand{\calV}{\mathcal{V}}
\newcommand{\calR}{\mathcal{R}}
\newcommand{\calX}{\mathcal{X}}
\newcommand{\calY}{\mathcal{Y}}
\newcommand{\calZ}{\mathcal{Z}}

\newcommand{\eps}{\epsilon}

\newcommand{\RCL}{\cR_{{\text{CL}}}}
\newcommand{\CPF}{\cC_{{\text{PF}}}}
\newcommand{\CIF}{\cC_{{\text{IF}}}}

\newcommand{\cC}{{\cal C}}
\newcommand{\cR}{{\cal R}}
\newcommand{\ok}[1]{#1}

\newcommand{\me}[1]{#1}

\title{Perfect vs. Independent Feedback in the Multiple-Access Channel}

\author{Oliver Kosut, Michelle Effros, Michael Langberg
\thanks{O. Kosut is with the School of Electrical, Computer and Energy Engineering at Arizona State University. Email: {\tt okosut@asu.edu}}
\thanks{M. Effros is with the Department of Electrical Engineering at the California Institute of Technology.
Email: \texttt{effros@caltech.edu}}
\thanks{M. Langberg is with the Department of Electrical Engineering at the University at Buffalo (State University of New York).  
Email: \texttt{mikel@buffalo.edu}}
\thanks{This work is supported in part by NSF grants CCF-1817241, CCF-1908725, and CCF-1909451. 
}
}

\begin{document}

\maketitle

\begin{abstract}
The multiple access channel (MAC) capacity with feedback is considered under feedback models designed to tease out which factors contribute to the MAC feedback capacity benefit.  Comparing the capacity of a MAC with ``perfect'' feedback, which causally delivers to the transmitters the true channel output, to that of a MAC with ``independent'' feedback, which causally delivers to the transmitters an independent instance of that same channel output, allows separation of effects like cooperation from alternative feedback benefits such as knowledge of the channel instance. Proving that the Cover-Leung (CL) achievability bound, which is known to be loose for some channels, is achievable also under (shared or distinct) independent feedback at the transmitters shows that the CL bound does not require transmitter knowledge of the channel instance.  Proving that each transmitter's maximal rate under independent feedback exceeds that under perfect feedback highlights the potential power of an independent look at the channel output. 
\end{abstract}

\section{Introduction}

While feedback does not increase the capacity of point-to-point memoryless channels \cite{shannon1956zero,dobrushin1958information}, it does significantly increase the capacity of distributed communication systems; examples where this occurs include the binary adder \me{\cite{gaarder1975capacity,king1978multiple,cover1981achievable} and Gaussian \cite{ozarow1984capacity} multiple-access channel (MAC)}.
Upper and lower bounds on MAC feedback-capacity, e.g., \cite{cover1981achievable,bross2005improved,venkataramanan2011new, hekstra1989dependence,kramer2006dependence,tandon2009outer},
are not tight in general but are known to be tight for specific MAC families \me{(e.g., \cite{willems1982feedback})}. 
A multi-letter capacity characterization of the MAC with feedback using the notion of directed information appears in \cite{kramer1998directed,kramer2003capacity}.
As noted by El Gamal and Kim in \cite{el2011network}, {\em ``feedback can enlarge the [MAC] capacity region by inducing statistical cooperation between the two senders.''}

Much like the information shared in MAC paradigms such as conferencing \cite{willems1983discrete}, cribbing \cite{willems1985discrete}, and the introduction of a cooperation facilitator \cite{noorzad2017unbounded},  feedback-information informs encoders about the messages of other encoders; this facilitates cooperation, and that cooperation has a significant impact on capacity in some channels.
However, unlike conferencing, cribbing, and the use of a cooperation facilitator, feedback can do more than just transfer information between MAC encoders; it also informs the encoders \me{about} the system realization, such as the channel noise or channel output.
Allowing  encoders access to the received symbol at the decoder allows MAC encoders to adaptively fine-tune the channel input to fit the current state of the decoder, an action we here call {\em innovation} since it resembles the concept of innovation in prediction (e.g., \cite{bode1950simplified}).
Such innovation lies, for example,   at the heart of the celebrated capacity achieving encoding scheme for the Gaussian MAC with feedback \cite{ozarow1984capacity}.

While the study of cooperation addresses the rate benefits obtained from MAC encoders that share common information in order to coordinate their channel inputs, \me{our study of innovation  addresses the capacity benefits stemming explicitly from the encoders' knowledge of system realizations}.
In order to distinguish the rate benefits of innovation from those of encoder cooperation in the MAC feedback, this work asks the following question.

\begin{question}
\label{q:main}
Can one isolate and analyze the impact of innovation on MAC feedback-capacity?
Specifically, is the fact that encoders have access to the system realization
crucial to rate gains obtained by feedback?
\end{question}




To answer Question~\ref{q:main}, for a given MAC, we  study the capacity region of two similar feedback models. 
The first is the standard MAC feedback model, in which the encoders have strictly causal access to the channel output received at the decoder. 
We call this traditional model the {\em perfect feedback} model.
The second is a modified version of feedback in which the encoders have strictly causal access to an outcome of the channel using a statistically identical and independent channel realization.
In this second version of feedback, called {\em independent feedback}, both encoders receive the same channel output, however that output differs from the one received at the decoder.\footnote{We also study a subtly different model, called {\em doubly-independent feedback} in which each encoder receives a different, independent copy of the channel output. All our results apply to both models.}
For example, consider the Gaussian MAC, in which  channel inputs $X_1$ and $X_2$ yield output $Y=X_1+X_2+Z$, where $Z$ has normal Gaussian distribution.
Here, in perfect feedback, the encoders receive $Y$; in  independent feedback, the encoders receive $Y'=X_1+X_2+Z'$, where $Z$ and $Z'$ are independent and identically distributed.
Notice that, given the independent feedback $Y'$, the encoders can no longer innovate as they do not gain any information about the system realization experienced by the decoder; however, they can still exchange information, enabling them to cooperate much as they would with perfect feedback. 
Thus, while traditional feedback can combine cooperation and innovation in an intertwined manner, our second notion of independent feedback filters out the impact of innovation, allowing us to explore  Question~\ref{q:main}.
\ok{Another motivation for this model is that it represents a situation where each transmitter receives a noisy version of the other transmitter's signal, and the noise at the transmitters is independent from that at the receiver. In fact, this phenomenon of each transmitter overhearing the other can arise naturally in practice, unlike perfect feedback which at best must be engineered.}
We seek to understand whether one form of feedback is superior to the other or whether, perhaps alternatively, the two models are incomparable.

In this work, we obtain the following results in the context of Question~\ref{q:main}. 
We first address the place of innovation in the classic rate region of Cover-Leung \cite{cover1981achievable} (here called the CL region and denoted by $\RCL$).
We show (in Theorem~\ref{thm:CL}) that any rate vector in the CL region, that is, any rate achievable using perfect feedback via \cite{cover1981achievable}, is also achievable using independent feedback; this implies that the CL region is achievable {\em without} innovation.
Denoting the perfect-feedback capacity by $\CPF$ and the independent-feedback capacity by $\CIF$, this proves that $\RCL \subseteq \CPF \cap \CIF$.
It follows, for example, that for any MAC in which the CL-region is tight (i.e., $\CPF=\RCL$) the perfect-feedback capacity is achievable using either independent feedback or perfect feedback, i.e., $\CPF=\RCL\subseteq \CIF$. 
One such ``tight'' family of MACs is the family of MACs in which, given the channel output, one user can derive the other's transmission  \cite{willems1982feedback}.
In this work, we extend (in Theorem~\ref{thm:erasure}) the family of MACs for which the CL-region is tight (thus $\CPF$ can be obtained without innovation) by enhancing known tight MAC families with an additional erasure component.

We next turn to the question of whether one form of feedback is superior to the other. 
In this context, we show that for some channels, the independent-feedback capacity region is not a subset of the perfect-feedback one, i.e., $\CIF \not \subset \CPF$. This shows, perhaps surprisingly, that for certain MACs there are rates achievable with independent 
feedback that are not achievable with perfect feedback.
In particular, we consider the \emph{single-rate capacities}---that is, the maximum achievable rate for one transmitter, not considering the other rate. By posing this problem as version of the relay channel, we show that for many channels, the single-rate capacities with independent feedback exceed those with perfect feedback. We present (in Theorem~\ref{thm:sufficient_condition}) a sufficient condition for this to occur, and for a sub-class of channels called \emph{additive channels} we provide (in Theorem~\ref{thm:additive}) a necessary and sufficient condition for it to occur.
This observation suggests a benefit from independent feedback that is neither cooperation (where encoders share knowledge of each other's messages) nor innovation (where encoders learn about the channel instance) but \emph{pooling}, where the encoders share what they learn from the feedback to effectively give the {\em decoder} two looks at the channel inputs. 

Our notion of independent feedback is a special case of {\em generalized feedback}, e.g., \cite{king1978multiple,carleial1982multiple,willems1983achievable,tang2007multiple,lapidoth2010awgn,lapidoth2012multiple,lapidoth2012multiple02,kramer2008topics}.
In generalized feedback, the 2-user MAC has three output channels $Y$, $Y_1$, $Y_2$, where $Y$ is the receiver information, $Y_1$ is the feedback information for user 1, and $Y_2$ is the feedback information for user 2. 
Prior results study  the case in which $Y_1$ and $Y_2$ describe channel state (or channel noise) information (e.g., \cite{lapidoth2012multiple,lapidoth2012multiple02}), the case in which $Y_1$ and $Y_2$ are noisy versions of $Y$ 
(e.g., \cite{king1978multiple,lapidoth2010awgn}), and other forms of non-perfect feedback (e.g., \cite{khisti2013multiple,shaviv2013multiple}).
An achievable region for generalized feedback is presented in \cite{carleial1982multiple}, through which the study of independent feedback for the $2$-user Gaussian MAC is presented as an example. 
To the best of our knowledge, the comparison between independent and perfect feedback and notions similar to MAC-innovation have not appeared previously in the literature.

\section{Model}

\emph{Notation:} For integers $n$, $m$, and $i$ we define $[n:m]=\{n,n+1,\ldots,m\}$ and $X^i=(X_1,X_2,\ldots,X_i)$. The robustly typical set \cite[Chapter 2]{el2011network} is denoted $T_\eps^{(n)}(X)$, where the distribution of $X$ is established through context. Jointly typical sets are usually denoted just by $T_\eps^{(n)}$, where the relevant variables and distribution are again established through context. Entropy and mutual information are denoted by standard notations $H(\cdot)$ and $I(\cdot;\cdot)$. Kullback-Leibler divergence is denoted by $D(p(x)\|q(x))$.

A \me{MAC} with two transmitters is given by a tuple $(\calX_1\times\calX_2,p(y|x_1,x_2),\calY)$, where $\calX_1,\calX_2$ are input alphabets, $p(y|x_1,x_2)$ is the conditional distribution describing the operation of the channel, and $\calY$ is the output alphabet. In this paper, we are interested in the following four different versions of the MAC model, defined formally later in this section. The models differ in what kind of feedback is available at each transmitter. We denote each by a two-letter abbreviation.
\begin{itemize}
\item No feedback (NF): The standard MAC model.
\item Perfect feedback (PF): Each transmitter receives the channel output exactly.
\item Independent feedback (IF): Each transmitter receives the same independent copy feedback output.
\item Doubly-independent feedback (DF): Each transmitter receives a different independent copy feedback output.
\end{itemize}
We define an $(n,R_1,R_2)$ code for each of these models as follows. For the NF model, the encoding function at transmitter $j$ at time $i$ is given by
\begin{equation}
f_{ji}:[1:2^{nR_j}]\to\calX_j
\end{equation}
whereas for the other three models it is
\begin{equation}
f_{ji}:[1:2^{nR_j}]\times \calY^{i-1}\to\calX_j.
\end{equation}
The decoding function is given by
\begin{equation}
g:\calY^n\to[1:2^{nR_1}]\times[1:2^{nR_2}].
\end{equation}
The two messages $M_1,M_2$ are chosen uniformly at random from $[1:2^{nR_1}]$ and $[1:2^{nR_2}]$, respectively. At time $i\in[1:n]$, for the NF model the channel inputs are $X_{ji}=f_{ji}(M_j)$, and for the other three models they are $X_{ji}=f_{ji}(M_j,Y_j^{i-1})$, where \me{$Y_{ji}$ is the feedback received at transmitters $j$ }. To describe the feedback under each model, let $Y_i,Y'_i,Y''_i$ be three different, statistically identical, versions of the channel output at time $i$; that is,
\begin{multline}
p_{Y_i,Y'_i,Y''_{i}|X_{1i},X_{2i}}(y_i,y'_i,y''_{i}|x_{1i},x_{2i})
\\=p(y_i|x_{1i},x_{2i})p(y'_i|x_{1i},x_{2i})p(y''_{i}|x_{1i},x_{2i}),
\end{multline}
where on the right-hand side, each function is the channel model. Thus, we have the Markov chain
\begin{multline}
(M_1,M_2,X_1^{i-1},X_2^{i-1},Y^{i-1},Y'^{i-1},Y''^{i-1})\\
\to(X_{1i},X_{2i})
\to (Y_i,Y'_i,Y''_{i}).
\end{multline}
The three feedback models can now be described by
\begin{align}
\text{PF: }&Y_{1i}=Y_{2i}=Y_i\label{PF_model}\\
\text{IF: }&Y_{1i}=Y_{2i}=Y'_i\label{IF_model}\\
\text{DF: }&Y_{1i}=Y'_i,\ Y_{2i}=Y''_i.\label{DF_model}
\end{align}
The message estimates are determined by $(\hat{M}_1,\hat{M}_2)=g(Y^n)$. Given a code, the probability of error is
\begin{equation}
P_e=\mathbb{P}((\hat{M}_1,\hat{M}_2)\ne (M_1,M_2)).
\end{equation}
A rate-pair $(R_1,R_2)$ is achievable if there exists a sequence of $(n,R_1,R_2)$ codes with probability of error going to $0$. Each capacity region $\calC_{\text{NF}},\calC_{\text{PF}},\calC_{\text{IF}},\calC_{\text{DF}}$ is the closure of the set of achievable rate-pairs for the corresponding channel model.

\section{The Cover-Leung (CL) region applies to independent feedback}

The Cover-Leung region $\calR_{\text{CL}}$ is the set of rate-pairs $(R_1,R_2)$ that satisfy
\begin{align}
R_1&\le I(X_1;Y|U,X_2),\\
R_2&\le I(X_2;Y|U,X_1),\\
R_1+R_2&\le I(X_1,X_2;Y)
\end{align}
for some $p(u)p(x_1|u)p(x_2|u)$. The proof that $\calR_{\text{CL}}\subset \calC_{\text{PF}}$ appears in \cite{cover1981achievable}. \ok{Intuitively, the auxiliary variable $U$ represents information that is known to both transmitters from the feedback signal.} The following theorem shows that this region is also achievable under each independent feedback model. This theorem is a special case of the achievable region for generalized feedback from \cite{carleial1982multiple}; \ml{we provide a direct proof of it in Appendix~\ref{appendix:CL}.}

\begin{theorem}\label{thm:CL}
$\calR_{\text{CL}}\subset\calC_{\text{IF}}$ and $\calR_{\text{CL}}\subset\calC_{\text{DF}}$.
\end{theorem}

\subsection{Extended channel families for which the CL-region is tight}

Given any MAC, we modify the channel model by appending an erasure channel, as follows. Let $W$ be a random variable with alphabet given by $\calY\cup\{\textsf{e}\}$, where $\textsf{e}$ is a symbol not contained in $\calY$. With probability $1-p$, $W=Y$, and with probability $p$, $W=\textsf{e}$ (i.e., an erasure). The choice of whether an erasure occurs is independent from all other randomness. We use $\calC_{\text{PF}}(W)$ to denote the capacity region with perfect feedback for the model with $W$ as the output, and $\calC_{\text{PF}}(Y)$ to denote the capacity region with perfect feedback with $Y$ as the output; $\calR_{\text{CL}}(W)$ and $\calR_{\text{CL}}(Y)$ designate the Cover-Leung regions.

\begin{theorem}\label{thm:erasure}
If $\calC_{\text{PF}}(Y)=\calR_{\text{CL}}(Y)$, then $\calC_{\text{PF}}(W)=\calR_{\text{CL}}(W)$.
\end{theorem}

\begin{IEEEproof}
\ml{See Appendix~\ref{appendix:erasure}.}
\end{IEEEproof}

\section{Independent feedback may be better than perfect feedback}
\label{sec:pf_if}

For a given MAC and for $j=1,2$, let $C_{j,\text{XY}}$ be the maximum achievable rate $R_j$ in feedback model XY; that is,
\begin{align}
C_{1,\text{XY}}&=\max\{R_1:(R_1,0)\in\calC_{\text{XY}}\}\\
C_{2,\text{XY}}&=\max\{R_2:(0,R_2)\in\calC_{\text{XY}}\}.
\end{align}
We show below that for many channels, independent feedback achieves higher single-rate capacities than perfect feedback. First we state the cut-set outer bound for this problem, which is a straightforward application of \cite[Theorem~18.4]{el2011network}.

\begin{proposition}
Let $\text{XY}\in\{\text{PF},\text{IF},\text{DF}\}$. If $(R_1,R_2)\in\calC_{\text{XY}}$, then there exists $p(x_1,x_2)$ such that
\begin{align}
R_1&\le I(X_1;Y,Y_2|X_2)\label{cutset_R1}\\
R_2&\le I(X_2;Y,Y_1|X_1)\\
R_1+R_2&\le I(X_1,X_2;Y),\label{cutset_sum_rate}
\end{align}
where the statistical relationship between $Y,Y_1,Y_2$ depends on the specific channel model via \eqref{PF_model}--\eqref{DF_model}.
\end{proposition}

Next we establish the single-rate capacities for the NF and PF models.
\begin{proposition}\label{prop:no_feedback}
For $(j,k)\in\{(1,2),(2,1)\}$,
\begin{equation}\label{single_capacity}
C_{j,\text{NF}}=C_{j,\text{PF}}=\max_{p(x_j),x_k} I(X_j;Y|X_k=x_k).
\end{equation}
\end{proposition}
\begin{IEEEproof}
Let $(j,k)=(1,2)$. (An analogous proof holds for $(j,k)=(2,1)$). Note that as $C_{1,\text{NF}}\le C_{1,\text{PF}}$, it is enough to prove achievability for the NF model, and the converse for PF. Achievability without feedback follows by simply sending the constant $X_{2i}=x_2$, and using a point-to-point code from transmitter 1. The converse for the PF model follows from \eqref{cutset_R1}. In particular, since $Y=Y_2$ in the PF model, if $(R_1,0)\in\calC_{\text{PF}}$, then 
\begin{align}
R_1\le \max_{p(x_1,x_2)} I(X_1;Y|X_2)
=\max_{p(x_1),x_2} I(X_1;Y|X_2=x_2).
\end{align}
\end{IEEEproof}

For the IF and DF models, certainly the single-rate capacities are at least that of the NF model. Thus, $C_{j,\text{IF}}$ and $C_{j,\text{DF}}$ are at least the quantity in \eqref{single_capacity}. The following theorem gives a sufficient condition \me{under which these capacities exceed \eqref{single_capacity}.}

\begin{theorem}\label{thm:sufficient_condition}
Let $(j,k)\in\{(1,2),(2,1)\}$. Suppose there exist $p^*(x_j)$, $x_k^*$ achieving the maximum in \eqref{single_capacity}, and  $\bar{x}_k\in\calX_k$ with 
\begin{align}
&I(X_j;Y|X_k=\bar{x}_k)+D(p_{Y|X_k}(y|\bar{x}_k)\|p_{Y|X_k}(y|x_k^*))\nonumber\\
&\cdot\left(1-\frac{H(Y|X_j,X_k=x_k^*)}{H(Y'|Y,X_k=x_k^*)}\right)>
I(X_j;Y|X_k=x_k^*)\label{sufficient_condition}
\end{align}
where $X_j\sim p^*(x_j)$. Then
\begin{equation}
C_{j,\text{IF}},C_{j,\text{DF}}>C_{j,\text{NF}}.
\end{equation}
\end{theorem}
\begin{IEEEproof}
Let $(j,k)=(1,2)$. Suppose there exist $p^*(x_1)$, $x^*_2$, $\bar{x}_2$ satisfying \eqref{sufficient_condition}.
We next show there is an achievable rate-pair $(R_1,0)$ where $R_1>C_{1,\text{NF}}$. The argument follows from the observation that \me{if we ignore the feedback at transmitter $1$, then} the independent feedback channel with $R_2=0$ is a relay channel. That is, transmitter 2 acts as a relay, with received signal $Y_2$. Since we ignore the feedback at transmitter 1, the statistical relationship between $Y_1$ and $Y_2$ does not matter, so the argument holds for both the IF and DF models, and we \me{can} write $Y'$ for the signal received at the relay. The compress-forward bound for the relay channel now gives the lower bound
\begin{align}
C_{1,\text{XY}}\ge \max_{\substack{p(x_1)p(x_2)\\ \cdot p(v|x_2,y')}} \min\{&I(X_1,X_2;Y)-I(Y';V|X_1,X_2,Y),\nonumber
\\ &I(X_1;V,Y|X_2)\}\label{cf_bound}
\end{align}
for $\text{XY}\in\{\text{IF},\text{DF}\}$. 

It remains to find $p(x_1)p(x_2)p(v|x_2,y_2)$ such that the above quantity exceeds $C_{1,\text{NF}}$. We let $p(x_1)=p^*(x_1)$. Let $\calV=\calY\cup\{\textsf{e}\}$, assuming $\textsf{e}$ is not an element of $\calY$. For parameters $a,b\in[0,1]$, define the following distributions:
\begin{align}
p(x_2)&=(1-a) 1(x_2=x_2^\star)+a\cdot 1(x_2=\bar{x}_2),\\
p(v|x_2,y')&=\begin{cases} b, & v=y'\\ 1-b, & v=\textsf{e}.\end{cases}
\end{align}
Note that
\begin{align}
I(Y';V|X_1,X_2,Y)&=b\, H(Y'|X_1,X_2,Y)
\\&=b\,H(Y|X_1,X_2),
\end{align}
where the second equality holds because $Y'$ is an independent copy of the channel output. 
Moreover
\begin{align}
&I(X_1;V,Y|X_2)=I(X_1;Y|X_2)+I(X_1;V|X_2,Y)
\\&=I(X_1;Y|X_2)+b\,I(X_1;Y'|X_2,Y).
\end{align}
Thus, if we maximize \eqref{cf_bound} over $b$, we achieve
\begin{align}
&\max_{b\in[0,1]} \min\{I(X_1,X_2;Y)-I(Y';V|X_1,X_2,Y),\nonumber
\\ &\qquad I(X_1;V,Y|X_2)\}
\\&=I(X_1;Y|X_2)+\max_{b\in[0,1]}
\min\{I(X_2;Y)\nonumber
\\&\qquad-b\,H(Y|X_1,X_2),\ b\,I(X_1;Y'|X_2,Y)\}.
\end{align}
The optimal choice of $b$ is
\begin{align}
b&=\min\left\{1,\frac{I(X_2;Y)}{H(Y|X_1,X_2)+I(X_1;Y'|X_2,Y)}\right\}
\\&=\min\left\{1,\frac{I(X_2;Y)}{H(Y'|X_2,Y)}\right\}.\label{optimal_b}
\end{align}
In the limit as $a\to 0$, $X_2$ becomes deterministically equal to $x_2^*$. Thus
\begin{align}
\lim_{a\to 0}I(X_2;Y)&=0,\\
\lim_{a\to 0}H(Y'|X_2,Y)&=H(Y'|X_2=x_2^*,Y).
\end{align}
In order for the sufficient condition \eqref{sufficient_condition} to hold, we must have $H(Y'|X_2=x_2^*,Y)>0$. Thus, the limiting value of $I(X_2;Y)$ is strictly smaller than that of $H(Y'|X_2,Y)$, which means that for sufficiently small $a$,  $I(X_2;Y)\le H(Y'|X_2,Y)$. Thus, assuming that $a$ is sufficiently small, the optimal $b$ is given by the ratio term in \eqref{optimal_b}. This gives the achieved rate
\begin{align}\label{achieved_rate}
I(X_1;Y|X_2)+I(X_2;Y)\left(1-\frac{H(Y|X_1,X_2)}{H(Y'|X_2,Y)}\right).
\end{align}
At $a=0$, this quantity becomes simply $C_{1,\text{NF}}$. Thus it is enough to prove that its derivative in $a$ at $a=0$ is positive. It is not hard to show that
\begin{equation}
\frac{\partial}{\partial p(x_2)} I(X_2;Y)=D(p(y|x_2)\| p(y))+\log e.
\end{equation}
Thus, differentiating with respect to $a$,
\begin{align}
&\frac{d}{da}I(X_2;Y)\Big|_{a=0}\nonumber
\\&=-D(p(y|x_2^*)\|p(y))+D(p(y|\bar{x}_2)\|p(y))\Big|_{a=0}
\\&=D(p(y|\bar{x}_2)\|p(y|x_2^*)).
\end{align}
Recall also that when $a=0$, $I(X_2;Y)=0$. In addition,
\begin{equation}
\frac{d}{da}I(X_1;Y|X_2)=-I(X_1;Y|X_2=x_2^*)+I(X_1;Y|X_2=\bar{x}_2).
\end{equation}
Putting this together, the derivative of \eqref{achieved_rate} with respect to $a$ at $a=0$ is
\begin{align}
&-I(X_1;Y|X_2=x_2^*)+I(X_1;Y|X_2=\bar{x}_2)\nonumber
\\&+D(p(y|\bar{x}_2)\|p(y|x_2^*))
\left(1-\frac{H(Y|X_1,X_2=x_2^*)}{H(Y'|Y,X_2=x_2^*)}\right).
\end{align}
Therefore, if \eqref{sufficient_condition} holds, the derivative of the achieved rate in \eqref{achieved_rate} with respect to $a$ is positive, so rates greater than $C_{1,\text{NF}}$ can be achieved.
\end{IEEEproof}

\section{Additive Channels}


Unfortunately, the sufficient condition in Theorem~\ref{thm:sufficient_condition} is sometimes difficult to verify, and, in general, we do not know if the condition is necessary. In this section, we study a sub-class of channels in which we can identify a necessary and sufficient condition for the single-rate capacities with independent feedback to exceed the single-rate capacities with no feedback or perfect feedback. In particular, many channels of interest are \emph{additive}, meaning that the channel behavior follows the operations of a group, as defined formally next.

\begin{definition}
    A MAC is \emph{additive} if there exists a group $\calG$ with operation $+$ and identity element $0$ where
    \begin{itemize}
        \item $\calX_1,\calX_2\subset\calG$,  and $0\in\calX_1\cap\calX_2$,
        \item the Markov chain $(X_1,X_2)\to Z\to Y$ holds where $Z=X_1+X_2$, and the alphabet of $Z$ is
\begin{equation}
\calZ=\{z\in\calG:x_1+x_2=z\text{ for some }x_1\in\calX_1,x_2\in\calX_2\},
\end{equation}
        \item there exists a function\footnote{It is with some abuse of notation that we call this function $+$, but it should be unambiguous.} $+:\calY\times\calG\to\calY$ such that, for any $g_1,g_2\in\calG$, $(y+g_1)+g_2=y+(g_1+g_2)$, $y+0=y$, and, for any $z,z'\in\calZ$,
\begin{equation}\label{additivity}
p_{Y|Z}(y|z)=p_{Y|Z}(y+(z'-z)|z').
\end{equation}
    \end{itemize}
\end{definition}


The following theorem gives necessary and sufficient conditions for the single-rate capacities with independent feedback to exceed those with perfect feedback in additive channels. The sufficient condition is derived from Theorem~\ref{thm:sufficient_condition}, and the necessary condition \me{follows} from the cut-set bound, each specialized using the algebraic structure of an additive channel.

\begin{theorem}\label{thm:additive}
Consider an additive MAC. Let $j\in\{1,2\}$. $C_{j,\text{IF}}=C_{j,\text{DF}}=C_{j,\text{NF}}$ if either of the following hold:
\begin{enumerate}
\item $\max_{p(x_1,x_2)} I(X_1,X_2;Y)=C_{j,\text{NF}}$.
\item For any $p(z)$ with support in $\calX_j$, there exists a random variable $K$ where $H(K|Z)=H(K|Y)=0$, and $Z\to K\to Y$ is a Markov chain.
\end{enumerate}
Conversely, if neither of the above \me{holds}, then $C_{j,\text{IF}},C_{j,\text{DF}}>C_{j,\text{NF}}$.
\end{theorem}

\begin{IEEEproof}
    \ml{See Appendix~\ref{appendix:additive}.}
\end{IEEEproof}

The following examples illustrate additive channels for which the conditions of Theorem~\ref{thm:additive} are or are not satisfied.

\begin{example} (Binary additive erasure MAC)
Let $\calX_1=\calX_2=\{0,1\}$, $\calY=\{0,1,2,\textsf{e}\}$. The channel is given by
\begin{equation}
p(y|x_1,x_2)=\begin{cases} 1-p, & y=x_1+x_2\\ p, & y=\textsf{e}\\ 0, & \text{otherwise}\end{cases}
\end{equation}
where $+$ denotes regular integer addition. This channel is additive with respect to the integer addition group, since we may take $Z=X_1+X_2$, and \eqref{additivity} is satisfied if we define
\begin{equation}
y+g=\begin{cases}y+g, & y\in\mathbb{Z},\\ \textsf{e}, & y=\textsf{e}.\end{cases}
\end{equation}
For this channel, for $j\in\{1,2\}$, $C_{j,\text{NF}}=1-p$, whereas $\max_{p(x_1,x_2)}I(X_1,X_2;Y)=(1-p)\log 3$. Thus, the first condition in the theorem holds iff $p=1$. The second condition holds iff $p\in\{0,1\}$. In particular, for any $0<p<1$, $C_{j,\text{IF}},C_{j,\text{DF}}>C_{j,\text{NF}}$. Moreover, for $p=0$ (i.e., without erasures), this channel satisfies the sufficient condition from \cite{willems1982feedback}, so $\calC_{\text{PF}}=\calR_{\text{CL}}$. By Theorem~\ref{thm:CL}, the same holds for any $p$. Therefore, for any $0<p<1$,  $\calC_{\text{PF}}$ is a strict subset of $\calC_{\text{IF}}\cap\calC_{\text{DF}}$.
\end{example}

\begin{example} (Binary symmetric MAC)
Let $\calX_1=\calX_2=\calY=\{0,1\}$. The channel is given by $Y=X_1\oplus X_2\oplus N$, where $N\sim\text{Ber}(p)$, and $\oplus$ denotes mod-2 addition. This channel is additive with respect to the mod-2 addition group, with $Z=X_1\oplus X_2$. However, condition 1 of the theorem holds, since for $j\in\{1,2\}$,
\begin{equation}
C_{j,\text{NF}}=\max_{p(x_1,x_2)}I(X_1,X_2;Y)=1-H(p)
\end{equation}
where $H(p)$ is the binary entropy function. (Condition 2 of the theorem holds iff $p=0$.) Thus $C_{j,\text{IF}}=C_{j,\text{DF}}=C_{j,\text{NF}}$.
\end{example}

\section{Conclusions and open problems}
Below, we list some problems left open in this work.
Question~\ref{q:main}, and, in particular, the result in Section~\ref{sec:pf_if}, compare the MAC capacity region with perfect and independent feedback. The original intuition of the authors was that perfect feedback would be superior to independent feedback, as the former holds the potential for innovation.
However, the results of Section~\ref{sec:pf_if} give examples for which this intuition is incorrect. The advantage obtained in Section~\ref{sec:pf_if} for independent feedback stems from the fact that an independent view of the channel output allows, for example, 
transmitter 2 to act like a relay, aiding the transmission of messages in cases where no such aid would be possible using perfect feedback.
%
Specifically, for independent feedback, we exhibit a tradeoff in rates between encoders implying that $\CIF \not \subset \CPF$.

A better understanding of the relationship between $\CIF$ and $\CPF$, and thus a better understanding of the answer to Question~\ref{q:main}, is left open in this work. Several questions arise naturally.
First, are there example MACs for which perfect feedback outperforms independent feedback ($\CPF \not \subset \CIF$), or, perhaps, is it always the case that $\CPF \subset \CIF$.
Is it the case, given a MAC, that either $\CPF \subset \CIF$ or $\CIF \subset \CPF$; or are there MACs for which both $\CIF \not \subset \CPF$ and $\CPF \not \subset \CIF$, \me{rending} the capacity regions incomparable.
The Gaussian MAC is an interesting example here: the perfect feedback capacity region was found in \cite{ozarow1984capacity}, which \me{uses} an achievable scheme that specifically makes use of feedback as \emph{innovation}. As such, we have not found a way to achieve the same rates with independent feedback.\footnote{However, for the Gaussian MAC, a version of the argument in Theorem~\ref{thm:sufficient_condition} can be used to show that the single-rate capacities for independent feedback exceed those of perfect feedback, simply because in the Gaussian relay channel, compress-forward outperforms direct transmission.}
It is also interesting to focus on the {\em sum-rate} and ask if there are example MACs for which \me{perfect feedback has a sum-rate advantage} over independent feedback (or, perhaps, vice-versa).
In fact, we have yet to find any example for which we can even prove that maximum achievable sum-rate differs between the two models.

\ok{Another open question has to do with the relationship between $\calC_{\text{IF}}$ and $\calC_{\text{DF}}$, which differ in that in the IF model both transmitters receive the same feedback signal, whereas in the DF model the transmitters receive different independent feedback signals. All of our results apply to both models equally, and so it is natural to ask whether their capacity regions could ever differ.}
These and other questions are a subject of future studies. 


\bibliographystyle{unsrt}
\bibliography{MAC_feedback}

\appendices

\section{Proof of Theorem~\ref{thm:CL}}\label{appendix:CL}

We give an achievability bound that applies for both the IF and DF models. 

\emph{Codebook generation:} Fix rates $R_0,R_1,R_2$, and distributions $p_U(u)p_{X_1|U}(x_1|u)p_{X_2|U}(x_2|u)$. For each $m_1\in[1:2^{nR_1}],m_2\in[1:2^{nR_2}]$, draw $m_0(m_1,m_2)$ uniformly at random from $[1:2^{nR_0}]$.  For each $m_0\in[1:2^{nR_0}]$, draw $u^n(m_0)\sim \prod_{i=1}^n p_U(u_i)$. For each $m_0,m_1,m_2$, draw
\begin{align}
x_1^n(m_0,m_1)\sim \prod_{i=1}^n p_{X_1|U}(x_{1i}|u_i(m_0)),\\
x_2^n(m_0,m_1)\sim \prod_{i=1}^n p_{X_2|U}(x_{2i}|u_i(m_0)).
\end{align}

\emph{Encoding:} Coding occurs over $B$ blocks, each of length $n$. For $j=1,2$, message $m_j=(m_{j,1},\ldots,m_{j,B-1})$ where $m_{j,b}\in[1:2^{nR_1}]$ for each $b\in[1:B-1]$. 
For notational ease, we also write $m_{j,B}=1$. Also let $m_{0,1}=1$ and $m_{0,b}=m_0(m_{1,b-1},m_{2,b-1})$ for $b\in[2:B]$. Prior to block $b$, we assume that $m_{0,b}$ is known to both transmitters, and $m_{0,b-1}$ is known to the receiver. (We prove below that this occurs with high probability.) In block $b$, transmitter $j$ sends $x_1^n(m_{0,b},m_{j,b})$.

\emph{Decoding:} Let $y_1^n(b)$ be the received vector at transmitter $1$ during block $b$. At the end of block $b$, transmitter $1$ finds the smallest $\hat{m}_{2,b}$ such that 
\begin{equation}(u^n(m_{0,b}),x_1^n(m_{0,b},m_{1,b}),x_2^n(m_{0,b},\hat{m}_{2,b}),y_1^n(b))\in T_\eps^{(n)}
\end{equation}
and then calculates $m_{0,b+1}=m_0(m_{1,b},\hat{m}_{2,b})$. Transmitter 2 decodes similarly. The decoder, having received vector $y^n(b)$ during block $b$, finds the smallest $\hat{m}_{0,b}$ for which $u^n(\hat{m}_{0,b},y^n(b))\in T_\eps^{(n)}$ and then finds $(\hat{m}_{1,b-1},\hat{m}_{2,b-1})$ such that $\hat{m}_{0,b}=m_0(\hat{m}_{1,b-1},\hat{m}_{2,b-1})$ and 
\begin{align}
&(u^n(m_{0,b-1}),x_1^n(m_{0,b-1},\hat{m}_{1,b-1}),x_2^n(m_{0,b-1},\hat{m}_{2,b-1}),y^n(b))\nonumber\\
&\hspace{2in}\in T_\eps^{(n)}.
\end{align}

\emph{Probability of error analysis:} By the packing lemma \cite[Lemma~3.1]{el2011network},  if $R_2<I(X_2;Y_1|X_1,U)$ then, after block $b$, transmitter $1$ decodes $m_{2,b}$ with high probability. Since, in each of IF and DF models, $Y_1$ is statistically identical to $Y$, we can write this condition as $R_2<I(X_2;Y|X_1,U)$. Similarly, transmitter $2$ decodes $m_{1,b}$ if $R_1<I(X_1;Y|X_2,U)$. If each transmitter decodes the other's message, then they can both compute $m_{0,b+1}$ with error.

After block $b$, if $R_0<I(U;Y)$, then the receiver decodes $m_{0,b}$ correctly with high probability. It then decodes $(m_{1,b-1},m_{2,b-1})$ correctly with high probability if
\begin{align}
R_1+R_2&<I(X_1,X_2;Y|U)+R_0,\\
R_1&<I(X_1;Y|U,X_2)+R_0,\\
R_2&<I(X_2;Y|U,X_1)+R_0.
\end{align}
We can therefore make $R_0$ arbitrarily close to $I(U;Y)$. The sum-rate condition is
\begin{align}
R_1+R_2&<I(X_1,X_2;Y|U)+I(U;Y)
\\&=I(U,X_1,X_2;Y)
=I(X_1,X_2;Y),
\end{align}
where we have used the Markov chain $U\to(X_1,X_2)\to Y$.

\section{Proof of Theorem~\ref{thm:erasure}}\label{appendix:erasure}

Since the Cover-Leung region is always achievable, we know that $\calR_{\text{CL}}(W)\subset\calC_{\text{PF}}(W)$. Thus it remains to prove that $\calC_{\text{PF}}(W)\subset \calR_{\text{CL}}(W)$. We first show that $\calR_{\text{CL}}(W)=(1-p)\calR_{\text{CL}}(Y)$. Let $E=\mathbf{1}(W=\textsf{e})$ be the indicator variable for an erasure occuring. Now, for any $p(u)p(x_1|u)p(x_2|u)$,
\begin{align}
&I(X_1;W|U,X_2)
\\&=I(X_1;W,E|U,X_2)
\\&=I(X_1;E|U,X_2)+I(X_1;W|U,X_2,E)
\\&=(1-p) I(X_1;W|U,X_2,E=0)+p I(X_1;W|U,X_2,E=1)
\\&=(1-p) I(X_1;Y|U,X_2,E=0)
\\&=(1-p) I(X_1;Y|U,X_2)
\end{align}
where we have used the facts that $E$ is independent of $(U,X_1,X_2,Y)$, and that if $E=0$ then $W=Y$. By similar arguments,
\begin{align}
    I(X_2;W|U,X_1)&=(1-p) I(X_2;Y|U,X_1)\\
 I(X_1,X_2;W)&=(1-p)I(X_1,X_2;Y).
\end{align}
This proves that $\calR_{\text{CL}}(W)=(1-p)\calR_{\text{CL}}(Y)$. Given the assumption that $\calC_{\text{PF}}(Y)=\calR_{\text{CL}}(Y)$, it remains to prove that $(1-p)\calC_{\text{PF}}(Y)\subset \calC_{\text{PF}}(W)$. Consider any $(R_1,R_2)\in\calC_{\text{PF}}(Y)$. We next show that $((1-p)R_1,(1-p)R_2)\in\calC_{\text{PF}}(W)$. Consider an $(n,R_1,R_2)$ perfect feedback code for the $Y$-channel with probability of error $P_e$. Fix an $\eps>0$. We next construct a code of length $n(1+\eps)/(1-p)$ for the $W$-channel with perfect feedback. The idea is to follow the code for the $Y$-channel, using the ``repeat if erasure'' strategy. That is, if $W\ne \textsf{e}$ (which, with perfect feedback, is known to the receiver and both transmitters), both transmitters proceed to transmit the next symbol according to the $Y$-channel code as they normally would. If $W=\textsf{e}$, then both transmitters repeat the same symbol again, until the received signal is not erased. This strategy only fails if fewer than $n$ non-erasures occur among the $n(1+\eps)/(1-p)$ received channel outputs. For sufficiently large $n$, the probability of this failure is arbitrarily small by the law of large numbers. Note that the number of message bits $nR_1,nR_2$ is unchanged from the $Y$-channel code. Thus, for any $\eps>0$, we have
\begin{equation}
    \left(\frac{(1-p)R_1}{1+\eps},\frac{(1-p)R_2}{1+\eps}\right)\in\calC_{\text{PF}}(W).
\end{equation}
Taking $\eps$ arbitrarily small completes the proof.


\section{Proof of Theorem~\ref{thm:additive}}\label{appendix:additive}

We begin by proving two lemmas pertaining to additive channels.

\begin{lemma}\label{lemma:permutation}
For an additive MAC, consider the conditional distribution $p_{Y|Z}$ as a probability matrix with each row representing a conditional distribution given $Z=z$. The rows of this matrix are permutations of each other.
\end{lemma}
\begin{IEEEproof}
Consider any $g\in\calG$. For any $y,y'\in\calY$, if $y'+g=y+g$, then
\begin{equation}
y'=y'+(g-g)=(y'+g)-g=(y+g)-g=y+(g-g)=y.
\end{equation}
Thus, $y'=y$ if and only if $y'+g=y+g$. In particular, $\calY+g=\calY$. In particular, for any $z,z'\in\calZ$, we have $\calY+(z'-z)=\calY$, which from \eqref{additivity} proves the lemma.
\end{IEEEproof}

\begin{lemma}\label{lemma:additive}
Consider an additive MAC. Let $(j,k)\in\{(1,2),(2,1)\}$. Given $p(x_j)$, $I(X_j;Y|X_k=x_k)$ does not depend on $x_k$.
\end{lemma}

\begin{IEEEproof}
Let $(j,k)=(1,2)$. Consider any $x_2,x_2'\in\calX_2$ where $x_2\ne x_2'$, and any $p(x_1)$. Let $Z=X_1+x_2$, let $\bar{Z}=X_1+x_2'$, let $Y$ be the output of the channel $p_{Y|Z}$ with $Z$ as its input, and let $\bar{Y}=Y+(x_2'-x_2)$. Note that the support of $\bar{Z}$ is contained in $\calX_1+x_2'$. For any $\bar{z}\in\calX_1+x_2'$ and any $\bar{y}$, we may write
\begin{align}
p_{\bar{Y}|\bar{Z}}(\bar{y}|\bar{z})
&=p_{Y|Z}(\bar{y}+(x_2-x_2')|\bar{z}+(x_2-x_2'))
\\&=p_{Y|Z}(\bar{y}|\bar{z})
\end{align}
 where in the second step we have used the fact that $\bar{z}\in\calX_1+x_2'\subset\calZ$, and $\bar{z}+(x_2-x_2')\in\calX+x_2\subset\calZ$,
so we can apply \eqref{additivity}. Thus, we have shown that $p_{\bar{Y}|\bar{Z}}=p_{Y|Z}$, so we may take $\bar{Y}$ to be the output of the channel with $\bar{Z}$ as the input. Moreover,
\begin{align}
I(X_1;Y|X_2=x_2)&=I(Z;Y)=I(\bar{Z};\bar{Y})
\\&=I(X_1;Y|X_2=x_2').
\end{align}
This proves the lemma.
\end{IEEEproof}

Now we prove Theorem~\ref{thm:additive}. Let $j=1$. Suppose the first condition in the theorem statement holds. Then, 
by the sum-rate condition in the cut-set bound \eqref{cutset_sum_rate}, for $\text{XY}\in\{\text{IF},\text{DF}\}$,
\begin{align}
C_{1,\text{XY}}\le \max_{p(x_1,x_2)} I(X_1,X_2;Y)
=C_{1,\text{NF}}.
\end{align}
This proves that $C_{1,\text{IF}}=C_{1,\text{DF}}=C_{1,\text{NF}}$.

Now suppose the second condition holds.
By the cut-set bound \eqref{cutset_R1}, for $\text{XY}\in\{\text{IF},\text{DF}\}$,
\begin{align}
C_{1,\text{XY}}&\le \max_{p(x_1,x_2)}I(X_1;Y,Y'|X_2) 
\\&=\max_{p(x_1),x_2} I(X_1;Y,Y'|X_2=x_2)
\\&=\max_{p(x_1)} I(X_1;Y,Y'|X_2=0)\label{second_cond1}
\\&=\max_{p(z):Z\in\calX_1} I(Z;Y,Y')\label{second_cond2}
\end{align}
where \eqref{second_cond1} follows from the additivity of the channel, and \eqref{second_cond2} follows because if $X_2=0$ then $Z=X_1$, which is supported on $\calX_1$. By assumption, for any $p(z)$ supported on $\calX_1$, there exists a random variable $K$ where $H(K|Z)=H(K|Y)=0$ and $Z\to K\to Y$ is a Markov chain. Since the joint statistics of $(Z,Y)$ are the same as $(Z,Y')$, we also have $H(K|Y')=0$. By the data processing inequality, $I(Z;Y)\le H(K)$, but also since $K$ is a deterministic function of both $Z$ and $Y$, we have $I(Z;Y)=I(Z,K;Y,K)\ge H(K)$. Thus $I(Z;Y)=H(K)$.
Now we may write
\begin{align}
    I(Z;Y,Y')&=I(Z;Y)+I(Z;Y'|Y)
    \\&=I(Z;Y)+I(Y,Z;Y')-I(Y;Y')
    \\&=I(Z;Y)+I(Z;Y')-I(Y;Y'|Z)-I(Y;Y')
    \\&=I(Z;Y)+I(Z;Y')-I(Y;Y')\label{second_cond3}
    \\&= 2H(K)-I(K,Y;K,Y')\label{second_cond4}
    \\&\le H(K)
    \\&=I(Z;Y)
\end{align}
where \eqref{second_cond3} follows because $Y\to Z\to Y'$ is a Markov chain, in \eqref{second_cond4} we have used the fact that $I(Z;Y)=I(Z;Y')=H(K)$ and that $K$ is a deterministic function of both $Y$ and $Y'$.
As the above holds for any $p(z)$ supported on $\calX$, 
\begin{align}
C_{1,\text{XY}}&\le \max_{p(z):Z\in\calX_1} I(Z;Y,Y')
\\&\le \max_{p(z):Z\in\calX_1} I(Z;Y)
\\&= \max_{p(x_1)} I(X_1;Y|X_2=0)\label{c1nf_derivation1}
\\&= \max_{p(x_1),x_2} I(X_1;Y|X_2=x_2)\label{c1nf_derivation2}
\\&= C_{1,\text{NF}}\label{c1nf_derivation3}
\end{align}
where \eqref{c1nf_derivation1} follows from the fact that if $X_2=0$, then $Z=X_1$, \eqref{c1nf_derivation2} follows from Lemma~\ref{lemma:additive}, and \eqref{c1nf_derivation3} follows from Proposition~\ref{prop:no_feedback}.
Therefore $C_{1,\text{IF}}=C_{1,\text{DF}}=C_{1,\text{NF}}$.

Now suppose that neither of the two conditions holds. We show that the sufficient condition of Thm.~\ref{thm:sufficient_condition} holds. Since the MAC is additive, there exists $p(x_1)$ where
\begin{equation}
I(X_1;Y|X_2=x_2)=C_{1,\text{NF}} \text{ for all }x_2\in\calX_2.
\end{equation}
If there is more than one distribution $p(x_1)$ achieving this maximum, take one with the largest possible support. This is well-defined, since the set of capacity-achieving input distributions is convex, so any distribution in the interior of this convex set will have the largest possible support.
The sufficient condition \eqref{sufficient_condition} holds if there exist $x_2^*,\bar{x}_2\in\calX_2$ where
\begin{align}
H(Y|X_1,X_2=x_2^*)&<H(Y'|Y,X_2=x_2^*)\label{entropy_condition}\\
D(p_{Y|X_2}(y|\bar{x}_2)\|p_{Y|X_2}(y|x^*_2))&>0.\label{KL_condition}
\end{align}
We claim $x_2^*=0$ satisfies \eqref{entropy_condition}. We proceed by contradiction, assuming that
\begin{equation}
H(Y|X_1,X_2=0)=H(Y'|Y,X_2=0).
\end{equation}
This implies that
\begin{equation}
I(X_1;Y'|Y,X_2=0)=0
\end{equation}
or equivalently
\begin{equation}
I(Z;Y'|Y)=0
\end{equation}
where $Z=X_1$. Thus, for all $z,y,y'$ where $p(z)>0$, $p(y|z)>0$, $p(y'|z)>0$,
\begin{equation}\label{y_prime_condition}
p(y'|y)=p(y'|z).
\end{equation}
For any $y$ let
\begin{equation}
r(y)=p_{Y'|Y}(y|y),
\end{equation}
so for any $z,y$ where $p(z)>0$ and $p(y|z)>0$,
\begin{equation}
    p(y|z)=r(y).
\end{equation}
Consider any $z,z',y$ where $p(z)>0$, $p(z')>0$, $p(y|z)>0$, and $p(y|z')>0$. For any $y'$ where $p(y'|z)>0$, we have
\begin{align}
r(y')&=p(y'|z)
\\&=p(y'|y)
\\&=\sum_{z} p_{Z|Y}(z|y) p_{Y|Z}(y'|z)
\\&=\sum_{z:p(y'|z)>0} p(z|y) r(y')
\\&=r(y')\, p_{Z|Y}(\{z:p(y'|z)>0\}|y).
\end{align}
Since $p(y'|z)>0$, we must have
\begin{equation}
    p_{Z|Y}(\{z:p(y'|z)>0\}|y)=1.
\end{equation}
In particular, if $z''$ is such that $p(z'')>0$ and $p(y'|z'')=0$, then
\begin{equation}
0=p(z''|y)=\frac{p(z'')p(y|z'')}{p(y)}
\end{equation}
so $p(y|z'')=0$. Since by assumption $p(y|z')>0$, we must have $p(y'|z')>0$. Now, applying \eqref{y_prime_condition} gives
\begin{equation}
p(y'|z)=p(y'|z')=r(y').
\end{equation}
As this holds for any $y'$ where $p(y'|z)>0$, in fact
\begin{equation}
p(y'|z)=p(y'|z')\text{ for all }y'\in\calY.
\end{equation}
To summarize the above, the set $\{z:p(z)>0\}$ can be divided into equivalence classes, where $z\sim z'$ if $p(y|z)>0,p(y|z')>0$ for some $y$. If $z\sim z'$, then the conditional distributions are equal: i.e., $p_{Y|Z=z}=p_{Y|Z=z'}$. As a consequence, it is clear that the optimal input distribution $p(z)$ puts equal probability mass on each equivalence class, and that $C_{1,\text{NF}}=I(Z;Y)=\log m$, where $m$ is the number of equivalence classes.  Let $z_1,z_2,\ldots,z_m$ be representative elements from each equivalence class, and let $\calY_k$ be the support of $p(y|z_k)$. By the above, $\calY_k$ are disjoint for different $k$. Since the optimizing distribution reaches all elements of $\calY$, it must be that $\bigcup_{k=1}^m \calY_k=\calY$. That is, $(\calY_1,\ldots,\calY_m)$ forms a partition of $\calY$. Note that
\begin{equation}\label{py_form}
    p(y)=\frac{p(y|z_k)}{m}\text{ if }y\in\calY_k.
\end{equation}
Recall from Lemma~\ref{lemma:permutation} that each $p(y|z)$ is a permutation of the others. Thus, it must be that $|\calY_k|=|\calY|/m$, and the non-zero elements of $p(y|z_k)$ are the same for each $k$.

If we define the random variable $K$ to be the index of the equivalence class of $Z$, then $H(K|Z)=0$. Since the $\calY_k$ sets are disjoint, also $H(K|Y)=0$. Because the conditional distribution $p_{Y|Z=z}$ is the same within an equivalence class for $z$, $Y\to K\to Z$ is a Markov chain. This very nearly proves that the second condition in the theorem holds, except that we have not proved this for any $p(z)$ supported on $\calX_1$. In particular, the $p(z)$ that we started with may not have full support.



Suppose there were some $z'\in\calX_1$ where $p(z')=0$. Since by assumption $p_Z$ maximizes $I(Z;Y)$ over all distributions with support in $\calX_1$, and does so with maximum support, this implies that if we had altered $p(z)$ to include any amount of $z'$, this would strictly decrease $I(Z;Y)$. In particular, let
\begin{equation}
    p_{a}(z)=(1-a)p(z)+a\,1(z=z').
\end{equation}
Let $I_a(Z;Y)$ be the mutual information where $Z\sim p_a$. In order for $p(z')=0$, we must have
\begin{equation}
    \frac{d}{da}I_a(Z;Y)\bigg|_{a=0}<0.
\end{equation}
We may calculate this derivative as follows:
\begin{align}
    &\frac{d}{da}I_a(Z;Y)\bigg|_{a=0}
    =\frac{d}{da} H_a(Y)\bigg|_{a=0}\label{derivative1}
    \\&=\sum_y (p(y)-p(y|z'))\log p(y)\label{derivative2}
    \\&=D(p(y|z')\|p(y))-H(Y)+H(Y|Z=z')
    \\&=D(p(y|z')\|p(y))-I(Z;Y)\label{derivative3}
    \\&=D(p(y|z')\|p(y))-\log m\label{derivative4}
\end{align}
where \eqref{derivative1} follows from the fact that $H(Y|Z=z)$ is identical for all $z\in\calZ$ by Lemma~\ref{lemma:permutation}, \eqref{derivative2} follows from a straightforward calculation, \eqref{derivative3} follows again by Lemma~\ref{lemma:permutation}, and \eqref{derivative4} follows from the above derivation that $I(Z;Y)=\log m$. Thus
\begin{align}
    0&>D(p(y|z')\|p(y))-\log m
    \\&=\sum_y p(y|z')\log \frac{p(y|z')}{m\, p(y)}
    \\&\ge -\log \left(\sum_{y:p(y|z')>0} m\, p(y)\right)\label{jensens}
\end{align}
where \eqref{jensens} follows from Jensen's inequality and the convexity of $-\log$. This implies
\begin{equation}
    \sum_{y:p(y|z')>0} m\, p(y)>1.
\end{equation}
Recall that, from \eqref{py_form}, $m\,p(y)=p(y|z_k)$ if $y\in\calY_k$. Thus, $m\,p(y)$ takes values among the non-zero elements of $p(y|z)$ (which are the same for each $z$). This implies that there is some $q>0$ where 
\begin{equation}\label{set_size_difference}
    |\{y:p(y|z')>0, m\,p(y)=q\}|>|\{y:p(y|z')=q\}|.
\end{equation}
Let $y_1,y_2,\ldots,y_r$ be an enumeration of the $y$ where $p(y|z')>0$ and $m\,p(y)=q$. For each $j\in[1:r]$ there is an $i_j$ where $y_j\in\calY_{i_j}$. In particular, $p(y_j|z_{i_j})=q$. Thus, by additivity,
\begin{equation}
    p_{Y|Z}(y_j+(z'-z_{i_j})|z')=q.
\end{equation}
Thus, $y_j+(z'-z_{i_j})\in\{y:p(y|z')=q\}$. But by \eqref{set_size_difference}, and the pigeon hole principle, there must be two $j\ne j'$ where 
\begin{equation}
y_j+(z'-z_{i_j})=y_{j'}+(z'-z_{i_{j'}}).
\end{equation}
Thus
\begin{equation}\label{pigeon_hole}
    y_j+(z_{i_{j'}}-z_{i_j})=y_{j'}.
\end{equation}
Let
\begin{equation}
    \bar{y}=y_j+(z_{i_{j'}}-z').
\end{equation}
Since $p(y_j|z')>0$, by additivity
\begin{equation}
    p_{Y|Z}(y_j+(z_{i_{j'}}-z')|z_{i_{j'}})>0
\end{equation}
and so 
\begin{equation}
    \bar{y}=y_j+(z_{i_{j'}}-z')\in\calY_{i_{j'}}.\label{contained_in_jprime}
\end{equation}
However, we may also write
\begin{align}
    \bar{y}
    &=y_j+(z_{i_{j'}}-z')
    \\&=y_j+(z_{i_{j'}}-z_{i_j})+(z_{i_j}-z')
    \\&=y_{j'}+(z_{i_j}-z')\label{applying_pigeon_hole}
\end{align}
where \eqref{applying_pigeon_hole} follows from \eqref{pigeon_hole}.
Since $p(y_{j'}|z')>0$, by additivity
\begin{equation}
    p_{Y|Z}(y_{j'}+(z_{i_j}-z')|z_{i_j})>0
\end{equation}
and so
\begin{equation}
    \bar{y}=y_{j'}+(z_{i_j}-z')\in\calY_{i_j}.\label{contained_in_j}
\end{equation}
Both \eqref{contained_in_jprime} and \eqref{contained_in_j} can only hold if $i_j=i_{j'}$, but by \eqref{pigeon_hole} this implies that $y_j=y_{j'}$. This is a contradiction since $j\ne j'$ and the $y_j$ were assumed to be distinct. This proves that, in fact, $p(z)>0$ for all $z\in\calX_1$, which implies the second condition in the theorem. Since by assumption this condition does not hold, we may conclude \eqref{entropy_condition} with $x_2^*=0$.



We next prove \eqref{KL_condition}, again using contradiction. Since we have taken $x_2^*=0$, we suppose that
\begin{equation}
\max_{\bar{x}_2\in\calX_2} D(p_{Y|X_2}(y|\bar{x}_2)\|p_{Y|X_2}(y|0))=0.
\end{equation}
This would imply that there exists $p(y)$ where, for all $x_2\in\calX_2$
\begin{equation}
p(y|x_2)=p(y).
\end{equation}
Moreover, we have that, for any $x_2\in\calX_2$,
\begin{equation}
C_{1,\text{NF}}=\max_{p(x_1)}I(X_1;Y|X_2=x_2)=\max_{p(z):Z\in(\calX_1+x_2)}I(Z;Y).
\end{equation}
By classical optimality conditions for the capacity formula \cite[Section~4.5]{gallager1968information}, there is a unique optimal output distribution $p(y)$ such that, for all $z\in(\calX_1+x_2)$,
\begin{equation}\label{capacity_optimality_condition}
D(p(y|z)\|p(y))\le C_{1,\text{NF}}.
\end{equation}
However, since $p(y)$ does not depend on the distribution of $X_2$, and $p(x_1)$ achieves $C_{1,\text{NF}}$ no matter what $x_2$ is, for exactly the same $p(y)$, \eqref{capacity_optimality_condition} holds for any
\begin{equation}
z\in\bigcup_{x_2\in\calX_2}(\calX_1+x_2)=\calZ.
\end{equation}
However, this would imply that
\begin{equation}
\max_{p(x_1,x_2)} I(X_1,X_2;Y)\le
\max_{p(z)} I(Z;Y)\le C_{1,\text{NF}}
\end{equation}
which would imply the first condition in the statement of the theorem. By the assumption that neither condition holds, this constitutes a contradiction, so we have proved \eqref{KL_condition}. 
%
%
%

\end{document}